\documentclass[11pt]{amsart}

\usepackage[utf8]{inputenc}
\usepackage{mathtools,amsthm,amssymb,amsfonts,mathrsfs, verbatim}
\usepackage{amsmath}	
\usepackage{typearea}
\usepackage[T1]{fontenc}
\usepackage{color}
\usepackage{mwe}
\usepackage{bbm}
\usepackage{listings}
\usepackage{etoolbox}
\usepackage{courier}
\usepackage{graphicx}
\usepackage{sidecap}
\usepackage{wrapfig}
\usepackage{enumerate}
 \usepackage{subfigure}

\usepackage{amsmath}
\usepackage{setspace}
\usepackage[tmargin=2.0cm, bmargin=2.0cm, lmargin=2.0cm, rmargin=2.0cm]{geometry}

\newtheorem{theorem}{Theorem}

\theoremstyle{definition}

\theoremstyle{definition}

\makeatletter
\def\maxwidth{ %
  \ifdim\Gin@nat@width>\linewidth
    \linewidth
  \else
    \Gin@nat@width
  \fi
}
\makeatother

\definecolor{fgcolor}{rgb}{0.345, 0.345, 0.345}

\usepackage{framed}
 {\par\unskip\endMakeFramed%
 \at@end@of@kframe}
\makeatother

\definecolor{shadecolor}{rgb}{.97, .97, .97}
\definecolor{messagecolor}{rgb}{0, 0, 0}
\definecolor{warningcolor}{rgb}{1, 0, 1}
\definecolor{errorcolor}{rgb}{1, 0, 0}

\title{Note on simulation pricing of $\pi$-options}

\author{Zbigniew Palmowski}
\address{Faculty of Pure and Applied Mathematics, Wroc\l aw University of Science and Technology, ul. Wyb. Wyspia\'nskiego 27, 50-370 Wroc\l aw, Poland}
\email{zbigniew.palmowski@gmail.com}

\author{Tomasz Serafin}
\address{Faculty of Pure and Applied Mathematics, Wroc\l aw University of Science and Technology, ul. Wyb. Wyspia\'nskiego 27, 50-370 Wroc\l aw, Poland}
\email{tomaszserafin.96@gmail.com}

\thanks{This work is partially supported by National Science Centre Grant No. 2016/23/B/HS4/00566 (2017-2020).}

\date{\today}

\begin{document}

\begin{abstract}
In this work, we adapt a Monte Carlo algorithm introduced by \cite{broadie_glasserman} to price a $\pi$-option.
		This method is based on the simulated price tree that comes from discretization and replication of possible trajectories
		of the underlying asset's price. As a result this algorithm produces the lower and the upper bounds that converge to the true price with the increasing depth of the tree.
		Under specific parametrization, this $\pi$-option is related to relative maximum drawdown and
		can be used in the real-market environment to protect a portfolio against volatile and unexpected price drops.
		We also provide some numerical analysis.

\vspace{3mm}

\end{abstract}

\maketitle

\pagestyle{myheadings} \markboth{\sc Z.\ Palmowski
--- T.\ Serafin} {\sc Pricing of watermark and $\pi$ options}

\vspace{1.8cm}


\newpage

\section{Introduction}
		In this paper we analyze $\pi$-options introduced by \cite{zervos} that depends on so-called relative drawdown
and can be used in hedging against volatile and unexpected price drops
or by speculators betting on falling prices. These options are the contracts with a payoff function:
		\begin{equation}
			g(S_T) =  ( M^{a}_T S^{b}_T - K)^{+}
			\label{eq:pi_call}
		\end{equation}
		\noindent in case of the call option and
		\begin{equation}
			g(S_T) =  ( K - M^{a}_T S^{b}_T )^{+}
			\label{eq:pi_put}
		\end{equation}
		\noindent in the case of put option, where
\begin{equation}\label{BS}
					S_t = S_0 \exp \left(\left(r -\frac{\sigma^2}{2}\right)t + \sigma B_t \right)
				\end{equation}
is an asset price in the Black-Scholes model under martingale measure, that is, $r$ is a risk-free interest rate, $\sigma$ is an asset's volatility and
$B_t$ is a Brownian motion. Moreover,
$$ M_t = \sup_{w \leq t} S_w $$
is a running maximum of the asset price and $T$ is its maturity.
Finally, $a$ and $b$ are some chosen parameters.
		
\begin{figure}[!h]
			\centering
			\includegraphics[width = .99\textwidth]{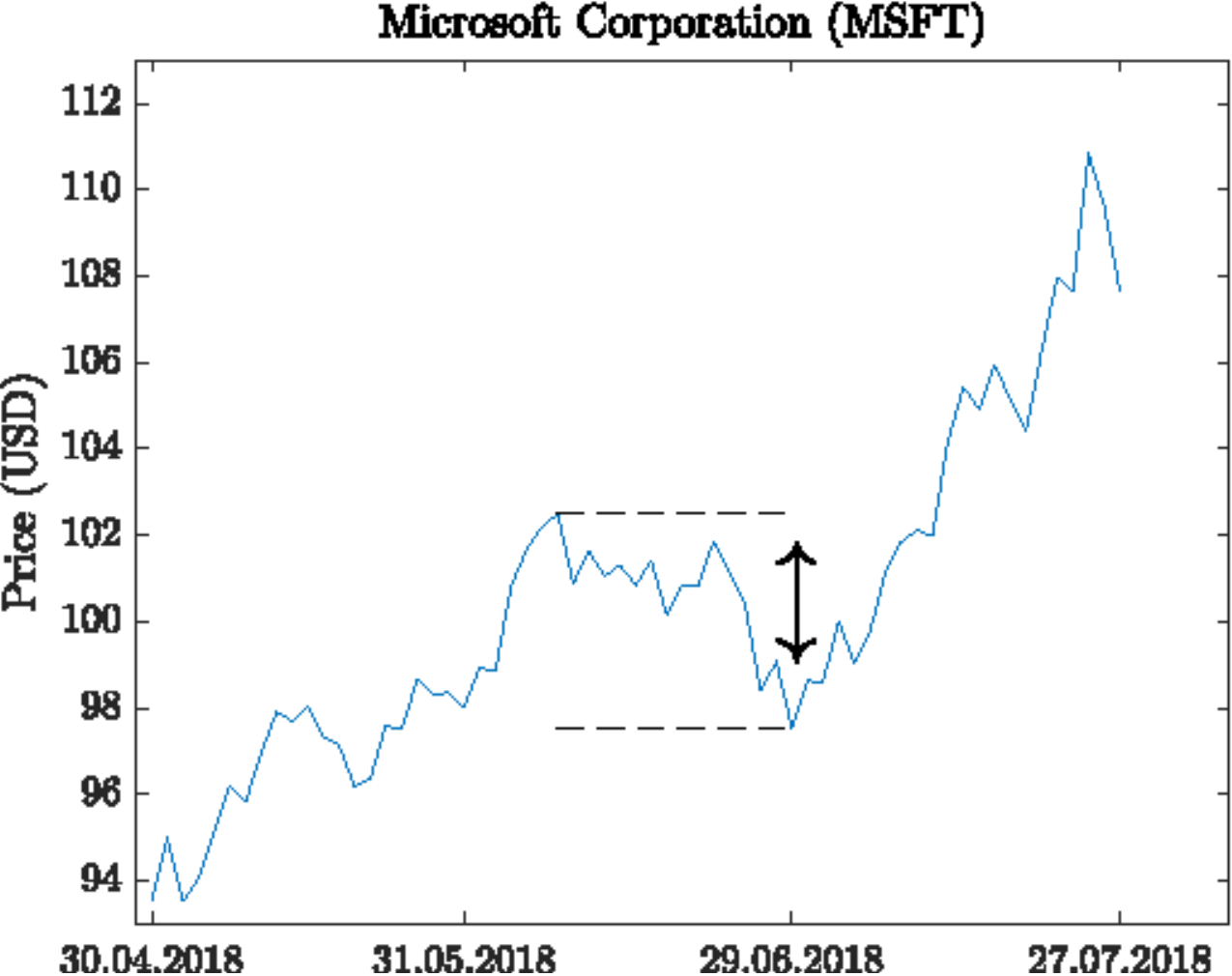}
			\caption{A sample drawdown for the Microsoft Corporation stock is marked with black arrows and dashed lines. Data is taken from \textit{www.finance.yahoo.com} }
			\label{fig:drawd}
		\end{figure}

Few very well known options are particular cases
of a $\pi$-option. In particular, taking $a=0$ and $b=1$
produces an American option and by choosing $a=1$ and $b=0$ we derive a lookback option.
Another interesting case, related to the concept of drawdown (see Figure \ref{fig:drawd}), is when
$-a = b = 1$ and $K=1$. Then the pay-out function
		$( K - M^{a}_T S^{b}_T )^{+}=1 - \frac{S^{b}_T}{M^{-a}_T} = \frac{M_T - S_t}{M_T} = D^{R}_T$ equals the relative drawdown $D^{R}_t$, defined as a quotient of the difference between maximum price and the present value of the asset and the past maximum price.
In other words, $D^{R}_t$ corresponds to the percentage drop in price from its maximum. We take a closer look at this specific parametrization of the $\pi$-option in the later sections, starting from Section \ref{ssec:piopt}.

Monte Carlo simulations are widely used in pricing in financial markets they have proved to be valuable and flexible computational tools to calculate the value of various options as witnessed by the contributions of \cite{1,2,3,4,5,6,7,8},\linebreak  \cite{9,10,11,11b},\linebreak \cite{12,13,13b,14,14b}, \linebreak \cite{5,16,17}.
One of the first attempts of Monte Carlo simulation for American options
is by \cite{15} where the backward induction algorithm was introduced. However, as appears later,
Tilley method suffers from exponentially increasing computational cost as the number of dimensions (assets) increases.
\cite{4}  to overcome this problem offered a non recombining binomial simulation approach instead
combined with some pruning technique to reduce computation burden and other variance reduction techniques to increase precision.
In the same year \cite{broadie_glasserman} construct
computationally cheap lower and upper bounds to the American option price. This method is used in this paper.
An alternative way to formulate the American
option pricing problem is in terms of optimal stopping
times. This is done in \cite{Carriere}, where it was proved that finding the price of American option
can be based on a backwards induction and calculating
a number of conditional expectations.
This observation gives another breakthrough in pricing early exercise derivatives by Monte Carlo
done by \cite{11b}.
They propose least square Monte Carlo (LSM) method which has proved to be versatile
and easy to implement.
The idea is to estimate the conditional expectation of the payoff from
continuing to keep the option alive at each possible
exercise point from a cross-sectional least squares
regression using the information in the simulated
paths. To do so we have to then solve some minimization problem.
Therefore this method is still computationally expensive.
Some improvements of this method have been also proposed; see also
\cite{Sten1, Sten2} who gave theoretical foundation of LSM and properties of its estimator.

There are other, various pricing methods in the case of American-type options; we refer
\cite{Zhao} for review. We have to note though that not all of them are good for simulation
of prices of general $\pi$-options as it is path-dependent product.
In particular, in pricing $\pi$-options one cannot use finite difference method introduced by \cite{Schw1, Schw2}
which uses a linear combination of the values of a function at three points
to approximate a linear combination of the values of derivatives of the same function at another point.
Similarly, the analytic method of lines of \cite{Carr}
is not available for pricing general $\pi$-options.
One can use though
a binomial tree algorithm (or trinomial model) though which goes backwards in time
by first discounting the price along each path and computing the continuation value.
Then this algorithm compares the
former with the latter values and decide for each path whether or not to exercise;
see \cite{Broadie} for details and references therein.
It is a common belief that Monte-Carlo method is more efficient
than binomial tree algorithm in case of path-dependent
financial instruments.
It has another known advantages as
handling time-varying variants, asymmetry, abnormal distribution and extreme conditions.

In this paper we adapt a Monte Carlo algorithm proposed in 1997 by \cite{broadie_glasserman} to price
		$\pi$-options. This numerical method replicates possible
		trajectories of the underlying asset's price by a simulated price-tree. Then, the values of two estimators, based on the price-tree, are obtained. They create an upper and a lower bound for the true price of the option and, under some additional conditions, converge to that price. The first estimator compares the early exercise payoff of the contract to its expected continuation value (based on the successor nodes) and decides if it is optimal to hold or to exercise the option. This estimation technique is one of the most popular ones used for pricing American-type derivatives. However, as shown by \cite{broadie_glasserman}, it overestimates the true price of the option. The second estimator also compares the expected continuation value and early exercise payoff, but in a slightly different way, which results in underestimation of the true price. Both Broadie-Glasserman Algorithms (BGAs) are explained and described precisely in Section \ref{sec:monte}. The price-tree that we need to generate is parameterized by the number of nodes and also by the number of branches in each node. Naturally, the bigger the numbers of nodes and branches, the more accurate price estimates we get. The obvious drawback of taking a bigger price-tree is that the computation time increases significantly with the size of the tree. However, in this paper we show that one can take a relatively small price-tree and still the results are satisfactory.

The Monte Carlo simulation presented in this paper
can be used in corporate finance and especially in portfolio management and personal finance planning.
Having American-type options in the portfolio, the analyst
might use the Monte Carlo simulation to determine its expected value even though
the allocated assets and options have varying degrees of risk, various correlations
and many parameters. In fact determining a return profile
is a key ingredient of building efficient portfolio.
As we show in this paper portfolio with $\pi$-options
out-performs typical portfolio with
American put options in hedging investment portfolio losses
since it allows investors to lock in profits
whenever stock prices reaches its new maximum.

In this paper we use BGA to price the $\pi$-option on relative drawdown for the Microsoft Corporation's (MSFT) stock and for the West Texas Intermediate (WTI) crude oil futures.
Input parameters for the algorithm are based on real market data.
Moreover, we provide an exemplary situation in which we explain the possible application of the $\pi$-option on relative drawdown to the protection against volatile price movements. We also compare this type of option to an American put and outline the difference between these two contracts.\\
		
\noindent This paper is organized as follows. In the next section we present the Broadie-Glasserman Algorithm.
In Section \ref{sec:numerical} we utilize this algorithm to numerically study $\pi$-options
for the Microsoft Corporation's stock and WTI futures.
Finally, in the last section, we state our conclusions and recommendations for further research in this new
and interesting topic.

	\section{Monte Carlo algorithm}	\label{sec:monte}
		
Formulas identifying the general price of $\pi$-option are known in some special cases and they are given in terms
of so-called scale functions and hence in terms of the solution of some second order ordinary differential equations; see
for example \cite[Chap. 5]{Chris} and \cite{Egami} for details and further references.
Still, the formulas are complex and a Monte Carlo method of pricing presented in this paper
is very efficient and accurate alternative method.
	In this section we present a detailed description of the used algorithm. In particular, we give formulas for two estimators,
one biased low and one biased high, that under certain conditions converge to the
		theoretical price of the option.\\
		
\subsection{Preliminary notations}				
\begin{figure}[!h]
			\centering
			\includegraphics[scale=0.8]{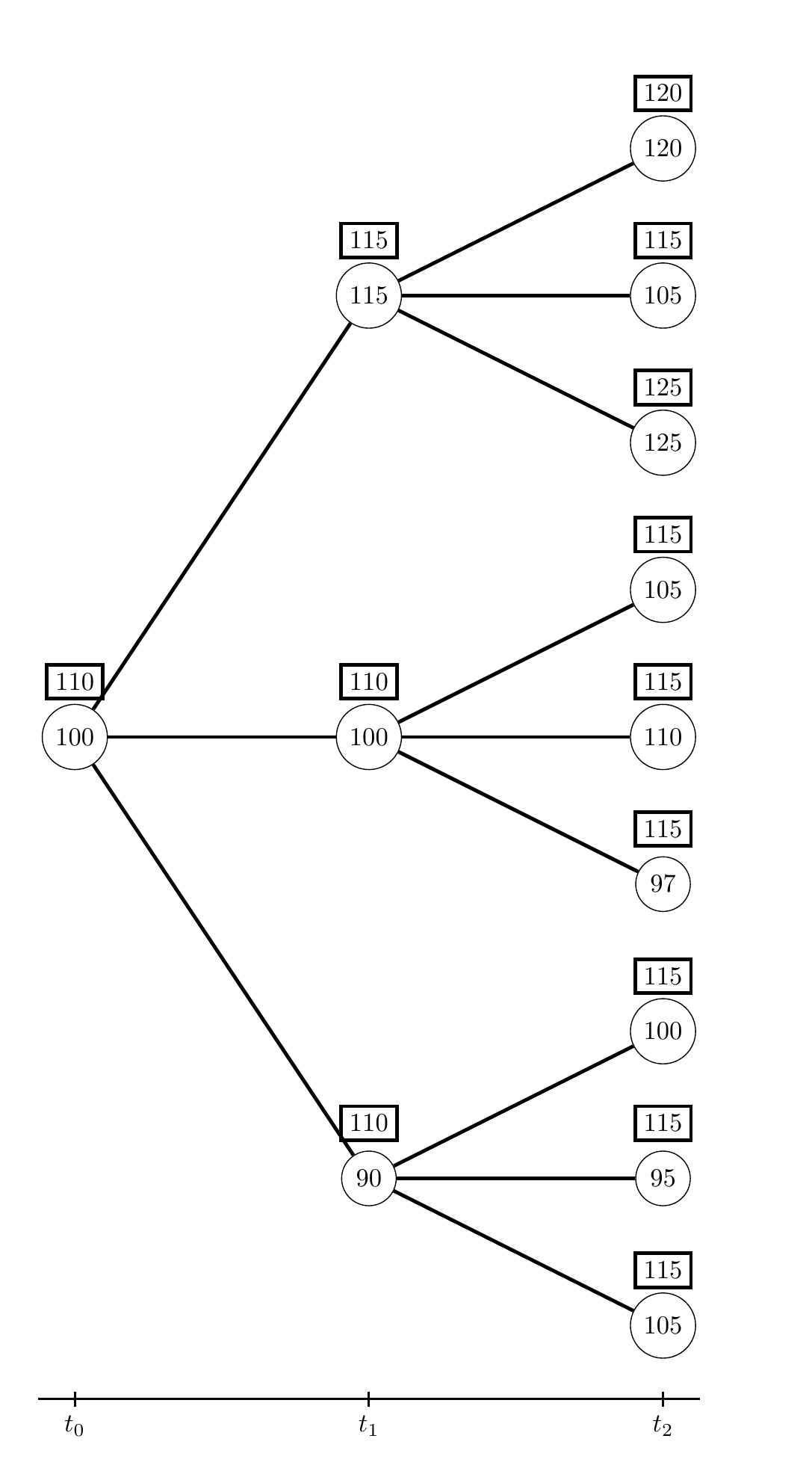}
			\caption{An example of the \textit{price-tree}. Underlying's price is marked with circles and corresponding maximums are marked with
			rectangles.}
			\label{fig:drzewko1}
		\end{figure}

		\noindent We adapt the Monte Carlo method
		introduced by \linebreak \cite{broadie_glasserman} for pricing American options.
		In this algorithm, values of two estimators are calculated on the so-called price-tree that represents the underlying's
		behavior over time. This tree is parametrized by the number of nodes $n$ and the number of branches in each node - denoted by $l$.
		For example the tree with parameters $n=2$, $l=3$ is depicted in Figure \ref{fig:drzewko1}.\\
		
\begin{figure}[!h]
			\centering
			\includegraphics[scale=0.8]{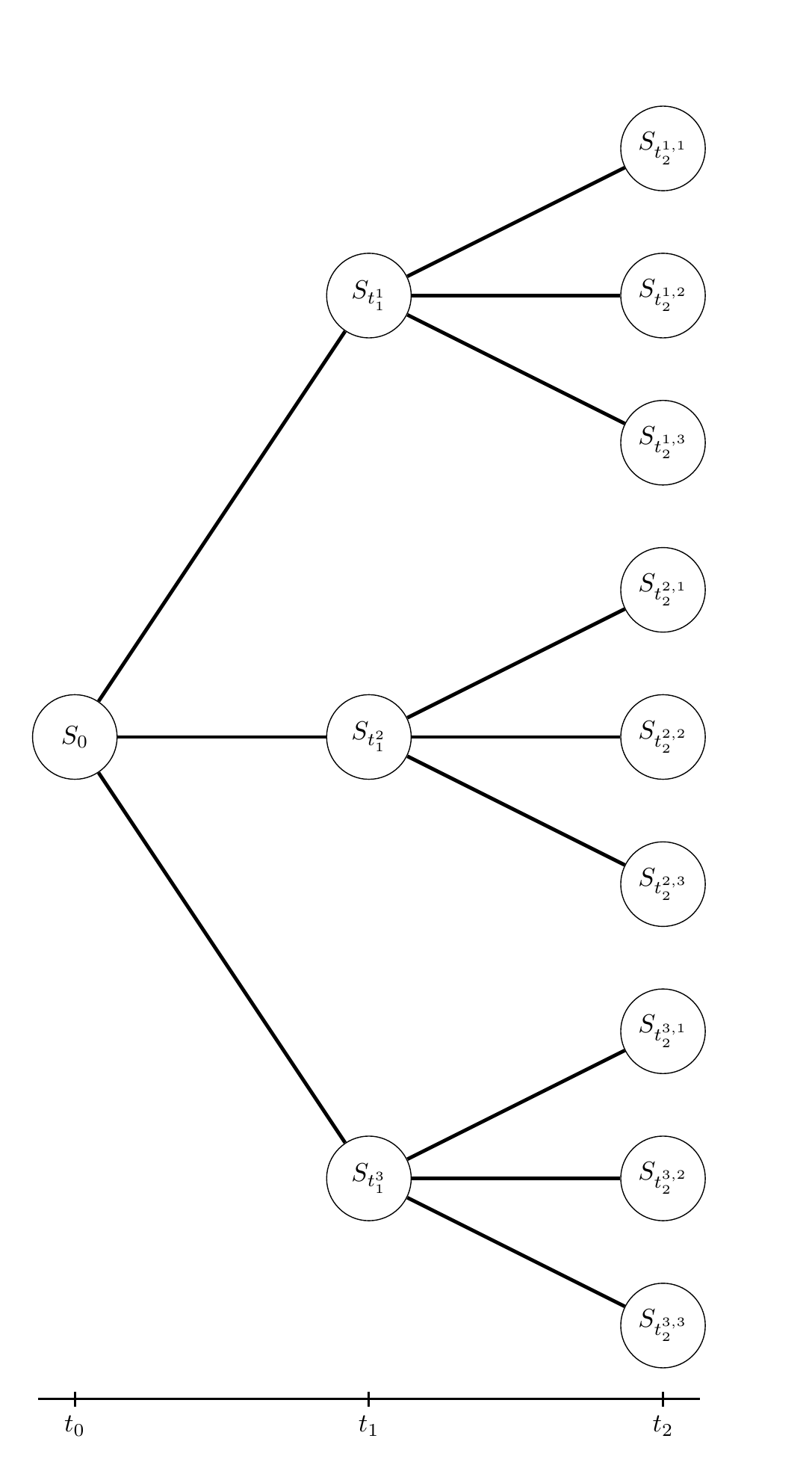}
			\caption{Branch selecting. }
			\label{fig:wyjasnienie_branches}
		\end{figure}

		\noindent In order to apply the numerical algorithm we have to discretize the price process given in \eqref{BS},
by considering the time sequence $t_0 = 0 <t_1<\ldots<t_n = T$ with $t_{i} = i \frac{T}{n}$ for $i=0,\ldots,n$.
By $$S_{t^{l_1,\ldots,l_i}_i}$$
		we denote the asset's price at the time $t_{i} = \frac{iT}{n}$. The upper index ${l_1,\ldots,l_i}$, associated with $t_i$,
		describes the branch selection (see Figure \ref{fig:wyjasnienie_branches}) in each of the tree nodes and allows us to uniquely
		determine the path of the underlying's price process up to time $t_i$.
Similarly, we define
		\begin{equation*}
		    M_{t^{l_1,\ldots,l_i}_i} = \max_{k \leq i} S_{t^{l_1,\ldots,l_k}_{k}}.
		\end{equation*}		
We introduce the state variable $\widetilde{S}_{t^{l_1,\ldots,l_i}_i} = (S_{t^{l_1,\ldots,l_i}_i}, M_{t^{l_1,\ldots,l_i}_i} )$ as well.		
		
We relate with it the payoff of an immediate exercise (for $\pi$-put) at time $t_i$ in the state $\widetilde{S}_{t^{l_1,\ldots,l_i}_i}$
given by
		 \begin{equation*}
				h_{t_i}(\widetilde{S}_{t^{l_1,\ldots,l_i}_i}) = (K-S^{a}_{t^{l_1,\ldots,l_i}_i}M^{b}_{t^{l_1,\ldots,l_i}_i})^{+}
		 \end{equation*}
and	the expected value of holding the option from $t_i$ to $t_{i+1}$, given asset's value
			$\widetilde{S}_{t^{l_1,\ldots,l_i}_i}$ at time $t_i$ defined via	
			\begin{equation*}
				g_{t_i}(\widetilde{S}_{t^{l_1,\ldots,l_i}_i}) = \mathbb{E} \left[ e^{-\frac{r}{n}} f_{t_{i+1}}(\widetilde{S}_{t^{l_1,\ldots,l_{i+1}}_{i+1}}) \middle\vert \widetilde{S}_{t^{l_1,\ldots,l_i}_i}  \right],
			\end{equation*}
where			
			\begin{equation*}
				f_{t_i}(\widetilde{S}_{t^{l_1,\ldots,l_i}_i}) = \max\{ h_{t_i}(\widetilde{S}_{t^{l_1,\ldots,l_i}_i}),  g_{t_i}(\widetilde{S}_{t^{l_1,\ldots,l_i}_i}) \}
			\end{equation*}
			 \noindent is the option value at time $t_i$ in state $\widetilde{S}_{t^{l_1,\ldots,l_i}_i}$.	
			Note that
			\begin{equation*}
				f_{t_n}(\widetilde{S}_{t^{l_1,\ldots,l_n}_n}) = f_{T}(\widetilde{S}_{T^{l_1,\ldots,l_n}})  = h_{T}(\widetilde{S}_{T^{l_1,\ldots,l_n}}) = (K-S^{a}_{T^{l_1,\ldots,l_n}}M^{b}_{T^{l_1,\ldots,l_n}})^{+}.
			\end{equation*}
		
\subsection{Estimators}
		
We will now give the formulas for the estimators $\Theta$ and $\Phi$ which overestimate and underestimate the true price of the option, respectively.
Then, we will state the main theorem showing that both estimators are asymptotically unbiased and that they converge to the theoretical price of the $\pi$-option. We also provide a detailed explanation of the estimation procedure based on the exemplary price tree. In all calculations we consider a $\pi$-put option with parameters $a=-1$, $b=1$ and $K=1$. Additionally we assume that the risk free rate used for discounting the payoffs equals $5\%$.
		
\subsubsection{The $\Theta$ estimator}
		
The formula for the estimator is recursive and given by:
				\begin{equation*}
			\Theta_{t_i} = \text{max}\left\{ h_{t_i}(\widetilde{S}_{t^{l_1,\ldots,l_i}_i}), e^{-\frac{r}{n}} \frac{1}{l} \sum_{j=1}^{l} \Theta_{t^{l_1,\ldots,l_i,j}_{i+1}}    \right\}, \hspace*{1cm} i=0,\ldots,n-1.
		\end{equation*}
At the option's maturity, $T$, the value of the estimator is given by
		\begin{equation*}
				 \Theta_{T} = f_{T}(\widetilde{S}_{T}).
		\end{equation*}

\begin{figure}[!h]%
			\centering
			\subfigure[Price-tree]{%
						\includegraphics[width=0.5\textwidth]{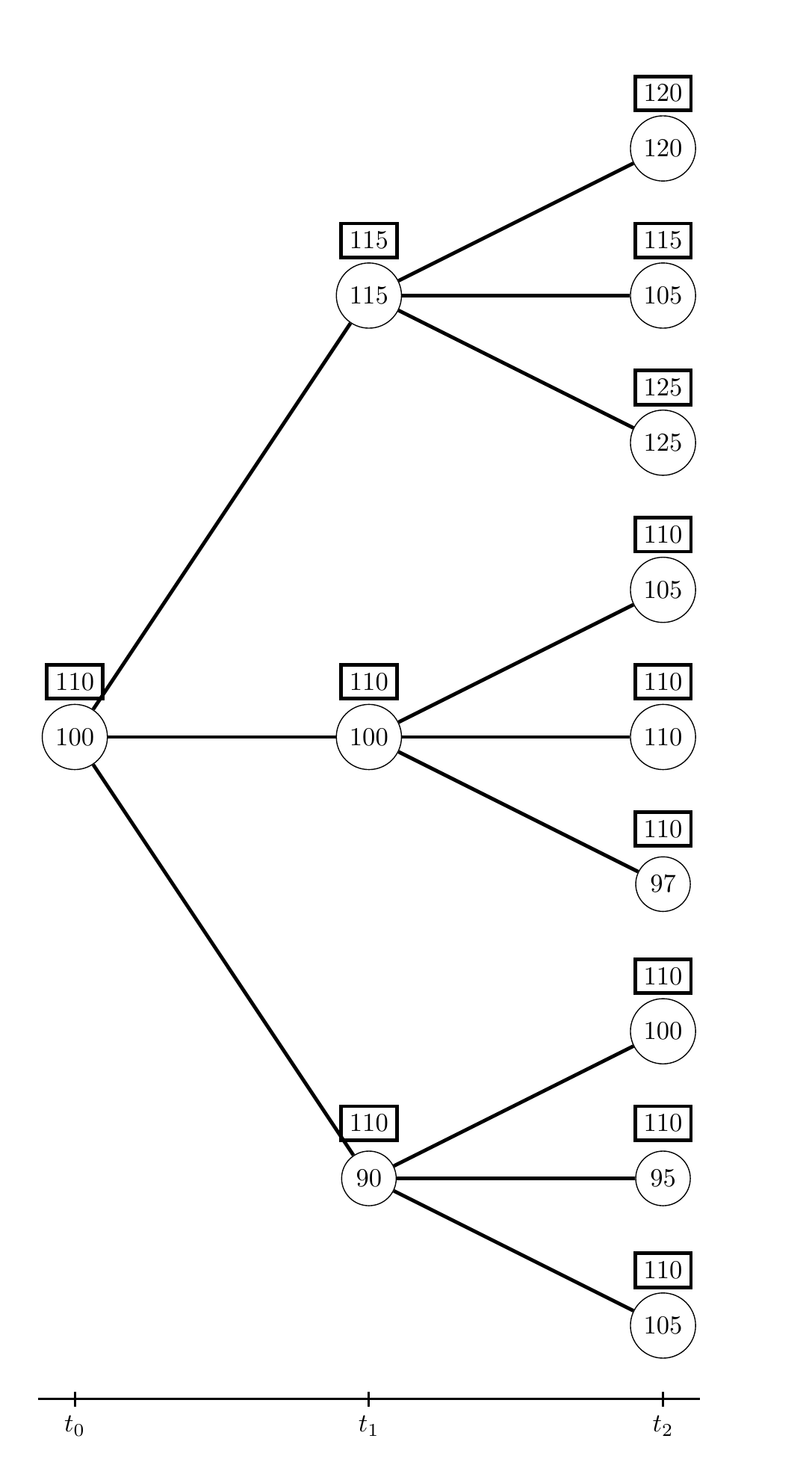}}%
									~~
			\subfigure[Evaluation of $\Theta$ estimator]{%
						\includegraphics[width=0.5\textwidth]{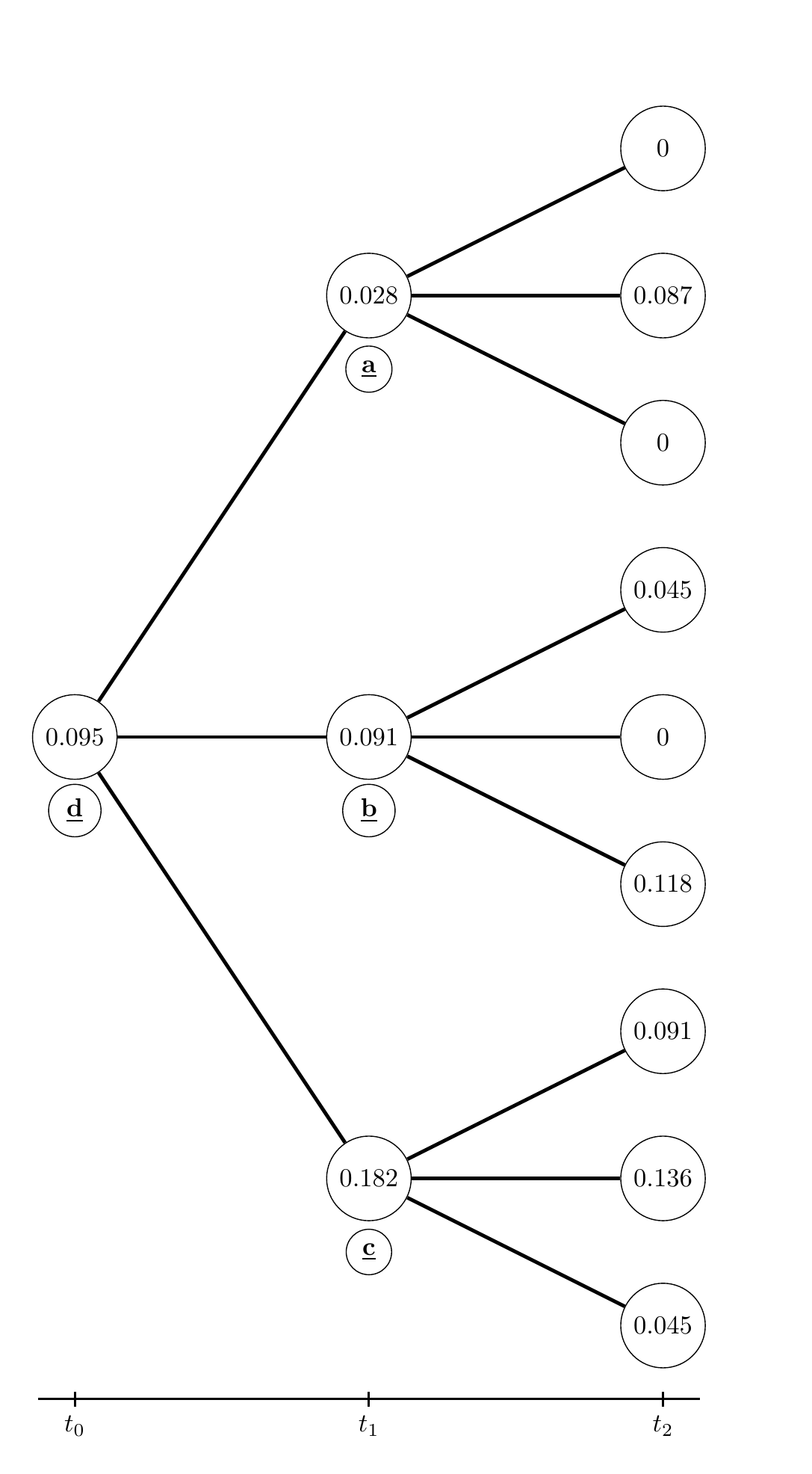}}%
			\caption{Explanation of $\Theta$ estimator }
			\label{fig:theta1}
		\end{figure}

		\noindent The $\Theta$ estimator, at each node of the price tree, chooses the maximum of the payoff of the option's early exercise at time $t_i$, $h_{t_i}(\widetilde{S}_{t^{l_1,\ldots,l_i}_i})$, and the expected continuation value, i.e. the discounted average payoff of successor nodes. Figure \ref{fig:theta1} shows how the value of $\Theta$ estimator is obtained given the certain realization of a price-tree.
		All calculations are also shown below:\\

	\begin{itemize}
		
		\item \textcircled{\underline{\textbf{a}}}	
$	\begin{cases}
			\text{Holding value: } \frac{0 + \frac{10}{115} + 0}{3} e^{-0.05} \approx \underline{0.028} \\
			\text{Early exercise: } 0
		\end{cases}$
		
		\item \textcircled{\underline{\textbf{b}}}
		$\begin{cases}
			\text{Holding value: } \frac{\frac{5}{110} + 0 + \frac{13}{110}}{3} e^{-0.05} \approx 0.052 \\
			\text{Early exercise: } \frac{10}{110} \approx \underline{0.091}
		\end{cases}$
		
		\item \textcircled{\underline{\textbf{c}}}
		$\begin{cases}
			\text{Holding value: } \frac{\frac{10}{110} +  \frac{15}{110} + \frac{5}{110}}{3} e^{-0.05} \approx 0.086 \\
			\text{Early exercise: } \frac{20}{110} \approx \underline{0.182}
		\end{cases}$
		
		\item \textcircled{\underline{\textbf{d}}}
		$\begin{cases}
			\text{Holding value: } \frac{0.028 + 0.091 + 0.182}{3} e^{-0.05} \approx \underline{0.095} \\
			\text{Early exercise: } \frac{10}{110} \approx 0.091
		\end{cases}$
		
		\end{itemize}

		\subsubsection{The $\Phi$ estimator}
		
		\noindent The $\Phi$ estimator is also defined recursively. Before we give the formula we need to introduce an auxiliary function $\xi$ by			
			\begin{equation} \label{eq:xi}				
				\xi^{j}_{t^{l_1,\ldots,l_i}_i} =  \begin{dcases}
											h_{t_i}(\widetilde{S}_{t^{l_1,\ldots,l_i}_i}), \text{ if } h_{t_i}(\widetilde{S}_{t^{l_1,\ldots,l_i}_i}) \geq e^{-\frac{r}{n}} \frac{1}{l-1} \sum_{\substack{k=1\\ k \neq j}}^{l} \Phi_{t^{l_1,\ldots,l_i,k}_{i+1}}\\
											e^{-\frac{r}{n}} \Phi_{t^{l_1,\ldots,l_i,j}_{i+1}}, \text{ if } h_{t_i}(\widetilde{S}_{t^{l_1,\ldots,l_i}_i}) < e^{-\frac{r}{n}} \frac{1}{l-1} \sum_{\substack{k=1\\ k \neq j}}^{l} \Phi_{t^{l_1,\ldots,l_i,k}_{i+1}}													
										\end{dcases}
			\end{equation}\\
			
			\noindent for $j=1,\ldots,l$. Now we can define the $\Phi$ estimator in the following way:
			
			\begin{equation}
				\begin{dcases}
						\Phi_{t^{l_1,\ldots,l_i}_i} = \frac{1}{l} \sum_{j=1}^{l} \xi^{j}_{t^{l_1,\ldots,l_i}_i}\\
						\Phi_{T} = f_{T}(\widetilde{S}_{T}).
				\end{dcases}
			\end{equation}

\begin{figure}[!h]%
			\centering
			\subfigure[Price-tree]{%
						\includegraphics[width=0.5\textwidth]{2.pdf}}%
									~~
			\subfigure[Evaluation of $\Phi$ estimator]{%
						\includegraphics[width=0.5\textwidth]{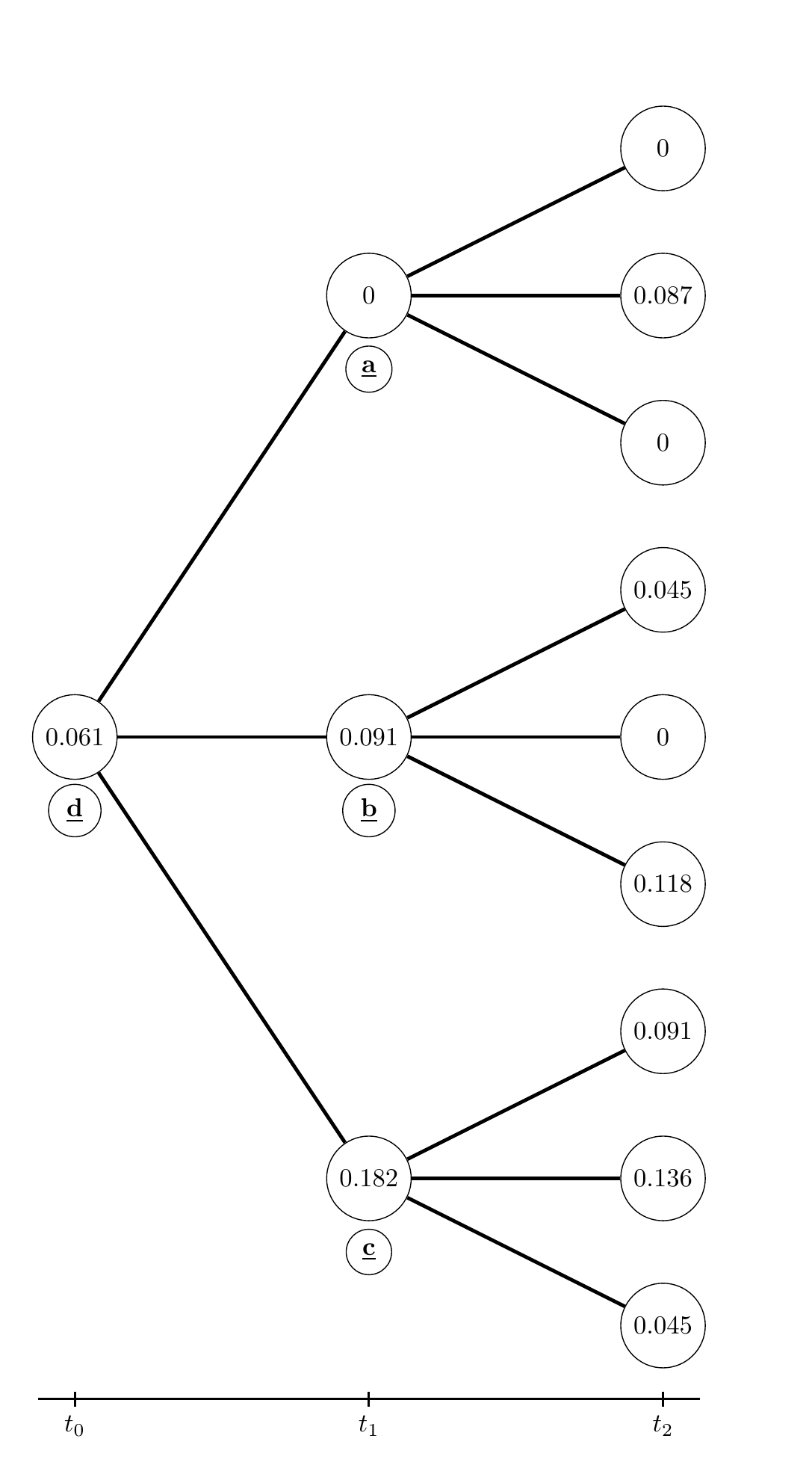}}%
			\caption{Explanation of $\Phi$ estimator}
			\label{fig:phi1}
		\end{figure}

		\noindent The formula for this estimator is more complicated. Therefore, we provide a detailed explanation of the mechanism behind the
		algorithm in the following part of this section. In our explanation we refer to Figure $\ref{fig:phi1}$. Note that in the following example, underlined numbers correspond to the final values associated with the specific branches of the tree.

		\begin{itemize}
		
		\item \textcircled{\underline{\textbf{a}}}	
		$\begin{cases}
			\text{Early exercise: } 0 \\
			\text{Holding value for branch $j=1$: } \frac{0.087+0}{2} e^{-0.05} \approx 0.041 > 0 \rightarrow \underline{0} \\
			\text{Holding value for branch $j=2$: } \frac{0 + 0}{2} e^{-0.05} = 0 \leq 0 = h_{i}(\widetilde{S}_{t_i}) \rightarrow \underline{0} \\			\text{Holding value for branch $j=3$: } \frac{0 + 0.087}{2} e^{-0.05} \approx 0.041 > 0 \rightarrow \underline{0} \\			
		\end{cases}$
		
	\noindent	For the branch $j=1$ we look at the two remaining ones to determine whether early exercising (payoff = 0) or holding the option (payoff = $\frac{0.087+0}{2} e^{-0.05}$) is more profitable. Obviously, early exercise is not optimal so we hold the option and thus, as the value of
		$\xi^{1}_{t^1_1}$ we take the payoff of the branch $j=1$ which is 0.\\
		
		\noindent For the branch $j=2$ both early exercise value and holding value from two other branches equals 0. Thus from $\eqref{eq:xi}$
		the value of $\xi^{2}_{t^1_1}$ equals the payoff of early exercise, which is 0.\\
		
		\noindent For the third branch, again holding the option is a more profitable decision (based on the payoffs of the two remaining
		branches). Thus $\xi^{3}_{t^1_1}$ takes the value corresponding to the branch $j=3$ and it is 0. \\
			
			\noindent Now the value of the estimator for node \textcircled{\underline{\textbf{a}}} is the sum of $\xi^{j}_{t^1_1}$ across all
			branches:
			$$\Phi_{t^1_1} = \frac{1}{3} \sum_{j=1}^{3} \xi^{j}_{t^1_1} = 0 . $$
		
		Similarly, we have the following values of our estimator.
		
		\item \textcircled{\underline{\textbf{b}}}
		$\begin{cases}
			\text{Early exercise: } \frac{10}{110} \approx \underline{0.091}\\
			\text{Holding value for branch $j=1$: } \frac{0 + 0.118}{2} e^{-0.05} \approx 0.056 < 0.091 \rightarrow \underline{0.091} \\
			\text{Holding value for branch $j=2$: } \frac{0.045 + 0.118}{2} e^{-0.05} \approx 0.078 < 0.091 \rightarrow \underline{0.091} \\			  \text{Holding value for branch $j=3$: } \frac{0.045 + 0}{2} e^{-0.05} \approx 0.021 < 0.091 \rightarrow \underline{0.091}

		\end{cases}$
		
		In this case, the value of the estimator for the \textcircled{\underline{\textbf{b}}} node equals 0.091.
		
		\item \textcircled{\underline{\textbf{c}}}
		$\begin{cases}
			\text{Early exercise: } \frac{20}{110} \approx \underline{0.182}\\
			\text{Holding value for branch $j=1$: } \frac{0.045 + 0.136}{2} e^{-0.05} \approx 0.086 < 0.182 \rightarrow \underline{0.182} \\
			\text{Holding value for branch $j=2$: } \frac{0.091 +0.045}{2} e^{-0.05} \approx 0.065 < 0.182 \rightarrow \underline{0.182} \\			   	\text{Holding value for branch $j=3$: } \frac{0.091 + 0.136}{2} e^{-0.05} \approx 0.108 < 0.182 \rightarrow \underline{0.182}
		\end{cases}$
		
		For node \textcircled{\underline{\textbf{c}}} the value of the estimator is 0.182.
		
		\item \textcircled{\underline{\textbf{d}}}
		$\begin{cases}
			\text{Early exercise: } \frac{10}{110} \approx \underline{0.091}\\
			\text{Holding value for branch $j=1$: } \frac{0.091 + 0.182}{2} e^{-0.05} \approx 0.13 > 0.091 \rightarrow \underline{0} \\
			\text{Holding value for branch $j=2$: } \frac{0.0 +0.182}{2} e^{-0.05} \approx 0.087 < 0.091 \rightarrow \underline{0.091} \\			   	\text{Holding value for branch $j=3$: } \frac{0 + 0.091}{2} e^{-0.05} \approx 0.043 < 0.091 \rightarrow \underline{0.091}
		\end{cases}$
		
		The value of the estimator for this node equals $\frac{0 + 0.091 + 0.091}{3} = 0.061$. This is also the (under)estimated value of the
		option.

		\end{itemize}

			
Following arguments of \cite{broadie_glasserman}, one can easily prove the following crucial fact.
			
			\begin{theorem}
				Both $\Theta$ and $\Phi$ are consistent and asymptotically unbiased estimators of the option value. They both converge to the true
				price of the option as the number of price-tree branches, $l$, increases to infinity. For a finite $l$:
				\begin{itemize}
					\item The bias of the $\Theta$ estimator is always positive, i.e.,
					$$ \mathbb{E}[ \Theta_0(l)] \geq f_0(\widetilde{S}_0). $$
					\item The bias of the $\Phi$ estimator is always negative, i.e.,
					$$ \mathbb{E}[ \Phi_0(l)] \leq f_0(\widetilde{S}_0). $$
				\end{itemize}	
				On every realization of the price-tree, the low estimator $\Phi$ is always less than or equal to the high estimator $\Theta$, i.e.,
				
				\begin{equation*}
				\mathbb{P} ( \Phi_{t^{l_1,\ldots,l_i}_i} \leq \Theta_{t^{l_1,\ldots,l_i}_i} ) = 1 .
				\end{equation*}
				
			\end{theorem}

			\section{Numerical analysis}\label{sec:numerical}
			
			In this section, we will present results of the numerical analysis. First, we use the algorithm described above to price the American
			option with arbitrary parameters.
			This will allow us to confirm that our Monte Carlo algorithm produces precise estimates of options' prices.
We focus on options related to Microsoft Corporation stock.
			Next, we price $\pi$-options for a number of combinations of parameters.
			We also consider $\pi$-option on drawdown using the real-market data and
			we compare it with an American put, which is one of the most popular tool for protecting our portfolio
against price drops.

\subsection{American options}
			
First of all we decided to check the robustness of the Monte Carlo pricing algorithm. We estimate prices of the American call options	with different strike prices. In the example, the uderlying asset price $S_0$ equals 100, $\sigma = 20\%$, risk free rate $r=5\%$ and the maturity is 30 days. In Table \ref{table:amer} we present the results of the estimation. Note that when utilizing the Broadie-Glasserman algorithm, we obtain the upper and the lower boundaries of the option price. To obtain the American option price estimate we average both values. 				
				\begin{table}[h!]
			\centering
				\begin{tabular}{|c|c|c|c|c|c|}
					\hline
					Strike & Low Est. & High Est. & Estimated Price & Real Price & Abs. Perc. Err. \\
					\hline
					\$80 & \$20.16 & \$20.55 & \$20.36 & \$20.33 &  0.14\% \\
					\hline
					\$85 & \$15.10 & \$15.54 & \$15.32 & \$15.35 &  0.19\%\\
					\hline
					\$90 & \$10.28 & \$10.62 & \$10.45 & \$10.43 & 0.19\% \\
					\hline
					\$95 & \$5.84 & \$6.02 & \$5.93 & \$5.89 &  0.68\% \\
					\hline
					\$100 & \$2.54 & \$2.60 & \$2.57 & \$2.51 &  2.36\% \\
					\hline
					\$105 & \$0.76 & \$0.77 & \$0.77 & \$0.73 &  5.33\% \\
					\hline
					\$110 & \$0.16 & \$0.16 & \$0.16 & \$0.14 &  13.35\% \\
					\hline
				\end{tabular}
\vspace{0.5cm}	

				\caption{Comparison of the estimated and 'real' American option prices with different strikes. Absolute percentage errors are also
				included
. }
				 \label{table:amer}
			\end{table}

	\subsection{$\pi$-options}\label{ssec:piopt}
			
\begin{figure}[!h]
			\centering
			\includegraphics[scale=0.9]{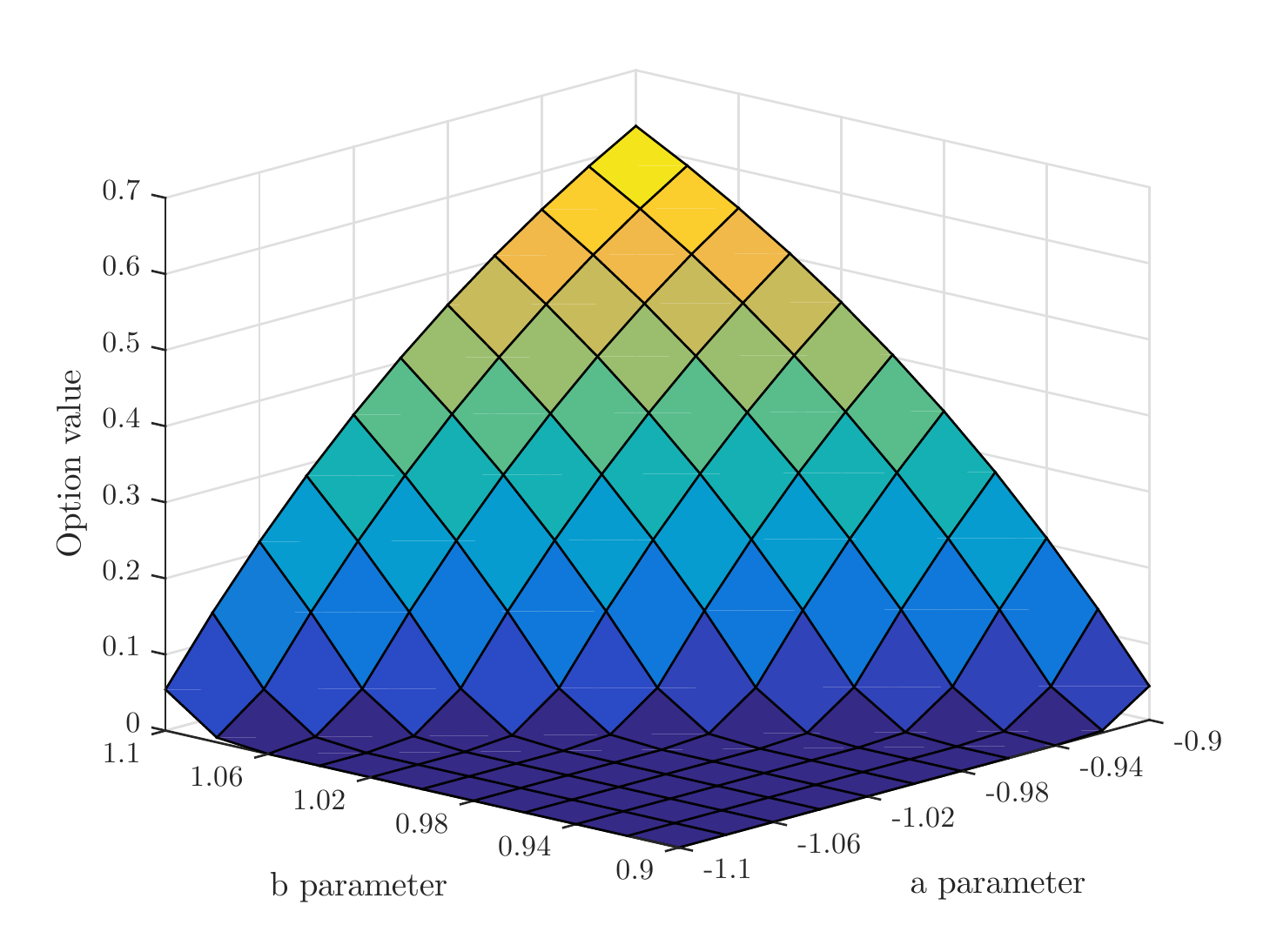}
			\caption{$\pi$-option price estimations for varying $a$ and $b$ - Microsoft Corporation stock.}
			\label{fig:drawdownparametry}
		\end{figure}

			We will analyze put $\pi$-option for various combinations of parameters $a$ and $b$.
We assume that parameter $a$ is varying from $-1.1$ to $-0.9$ and $b$ parameter between $0.9$ and $1.1$. The ranges of these parameters have been chosen arbitrarily for illustrative purposes.
			All input parameters for options pricing, $S_0$, $M_0$, volatility and interest rate are taken from the real-market data for the Microsoft Corporation stock (MSFT) and are
			 given in Table \ref{table:1}. The numerical results are presented in Figure \ref{fig:drawdownparametry}.

			\begin{table}[h!]
			\centering
				\begin{tabular}{|c|c|}
					\hline
					Parameter & Value \\
					\hline
					$S_0$ & 106.08 \\
					\hline
					$M_0$ & 110.83 \\
					\hline
					$\sigma$ & 17.03 \% \\
					\hline
					$r$ & 1.5\%\\
					\hline
					$l$ & 65 \\
					\hline
					$K$ & 1 \\
					\hline
				\end{tabular}
\vspace{0.5cm}	

				\caption{Input parameters for pricing $\pi$-put option on the Microsoft Corporation stock.}
				 \label{table:1}
			\end{table}

\subsection{$\pi$-options on relative drawdown}
\begin{figure}[!h]
			\centering
			\includegraphics[scale=0.9]{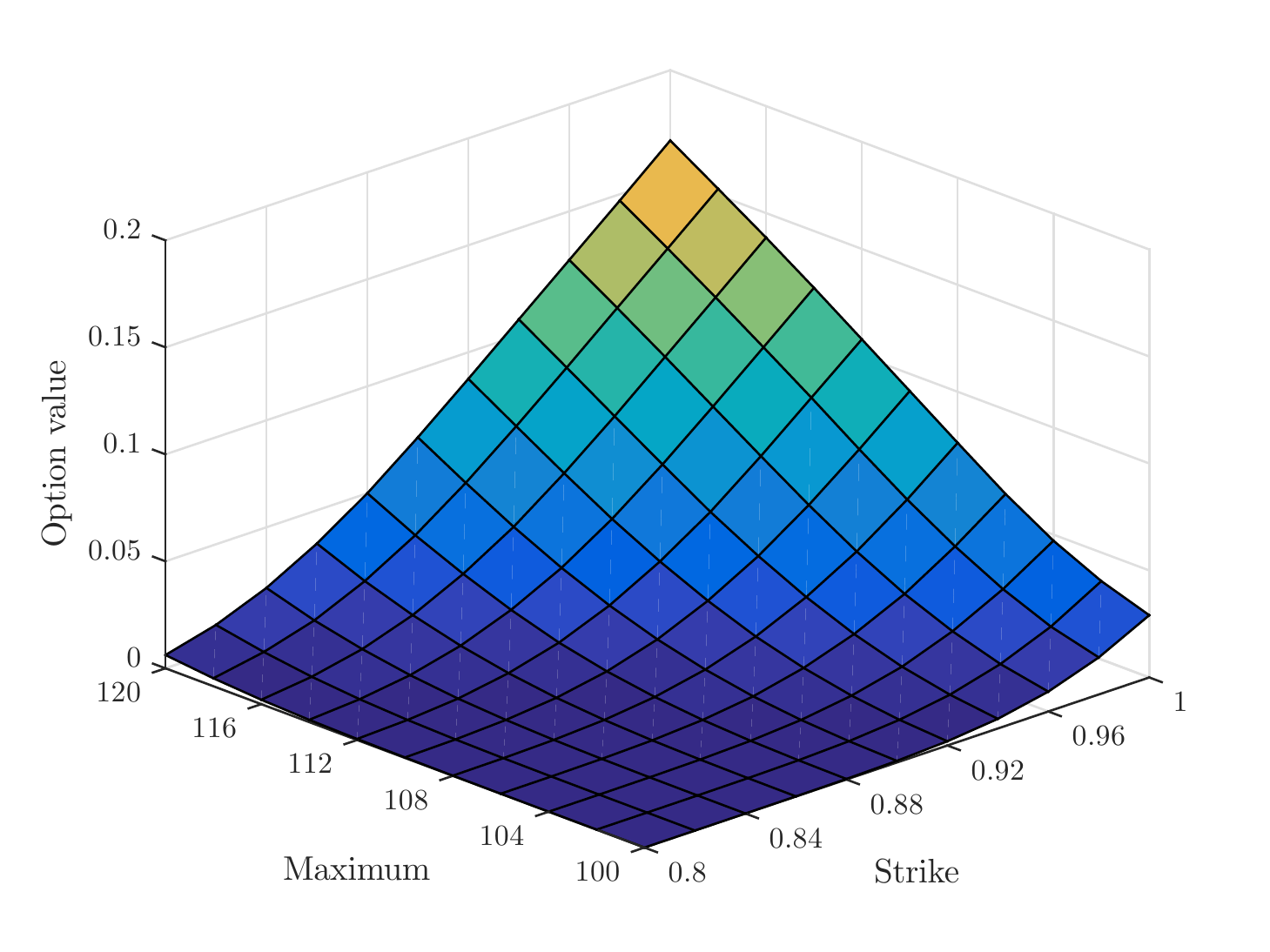}
			\caption{$\pi$-option on relative drawdown price estimations for varying $K$ and $M_0$ parameters. }
			\label{fig:drawdownpi}
		\end{figure}

		\noindent Recall that for $a=-1$ and $b=1$ the payoff of the $\pi$-option equals
			\begin{equation} \label{eq:relativedrawdown}
				\left( K - \frac{S_t}{M_t}  \right)^{ + },
			\end{equation}
		where $S_t/M_t$ is the current value of the relative drawdown of the underlying asset. We believe that such contracts could be very
		efficiently used for hedging and managing portfolio risk against the volatile drops in underlying's price (see Section \ref{ssec:pi_application}).
		One can adjust the payoff function \eqref{eq:relativedrawdown} by the appropriate choice of the
		strike $K$. The choice is arbitrary and solely dependent on the risk management goals of the option’s buyer. It allows to set the minimal size of drawdown we would like to protect against and let the buyer adjust and fully control the level of our exposure at risk associated with unexpected price drops. For example by setting $K=\frac{9}{10}$, the payoff of our option becomes greater than zero only if the drop in the price of the underlying from its maximum exceeds $10\%$. Of course, the bigger the value of $K$, the more expensive the option is.\\
	 We take a closer look at the impact of $M_t$ and $K$ on the price of this special case of $\pi$-option. Here, we assume that the maximum price $M_t$ is between $100$ and $120$ and $K$
		ranges between $0.8$ and $1$. This time, the remaining parameters, namely $S_0$, $r$ and $\sigma$, have been arbitrarily chosen for illustrative purposes and are given in Table \ref{table:2}. The results are shown in Figure \ref{fig:drawdownpi}.		
			
			\begin{table}[h!]
			\centering
				\begin{tabular}{|c|c|}
					\hline
					Parameter & Value \\
					\hline
					$S_0$ & 100 \\
					\hline
					$\sigma$ & 20 \% \\
					\hline
					$r$ & 5\%\\
					\hline
					$l$ & 100 \\
					\hline
				\end{tabular}
\vspace{0.5cm}	

				\caption{Input parameters for pricing $\pi$-option on relative drawdown.}
				 \label{table:2}
			\end{table}

\subsection{$\pi$-options on relative drawdown - application}	\label{ssec:pi_application}	
\begin{figure}[!h]
			\centering
			\includegraphics[width = .99\textwidth]{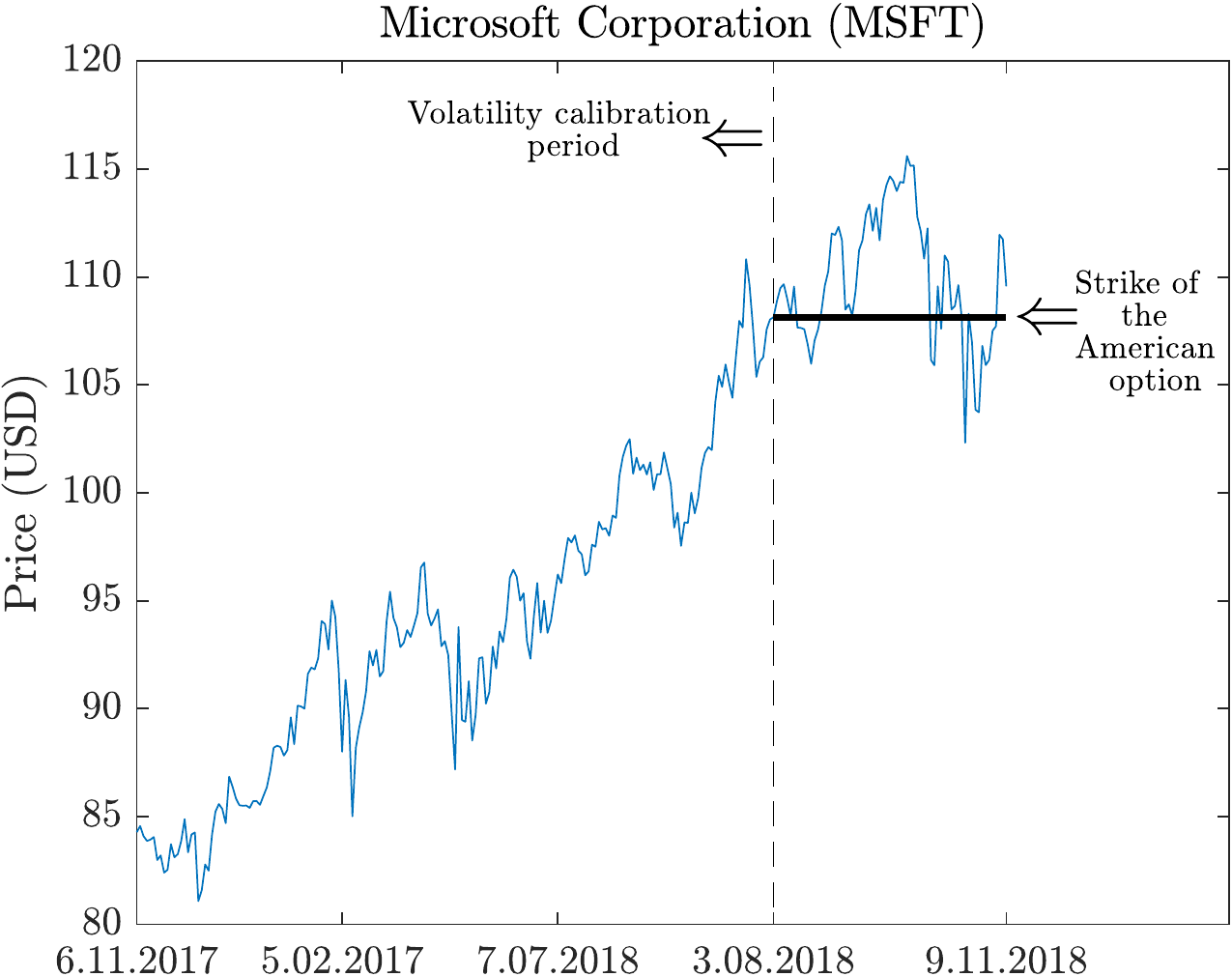}
			\caption{Daily closing prices of the Microsoft Corporation's stock. Data spans from 6.11.2017 to 9.11.2018. Vertical dashed line indicates the end of volatility calibration period. Option prices are calculated based on the volatility and the stock's price on 3.08.2018. }
			\label{fig:pi_kalibracja}
		\end{figure}

\begin{figure}[!h]
			\centering
			\includegraphics[width = .99\textwidth]{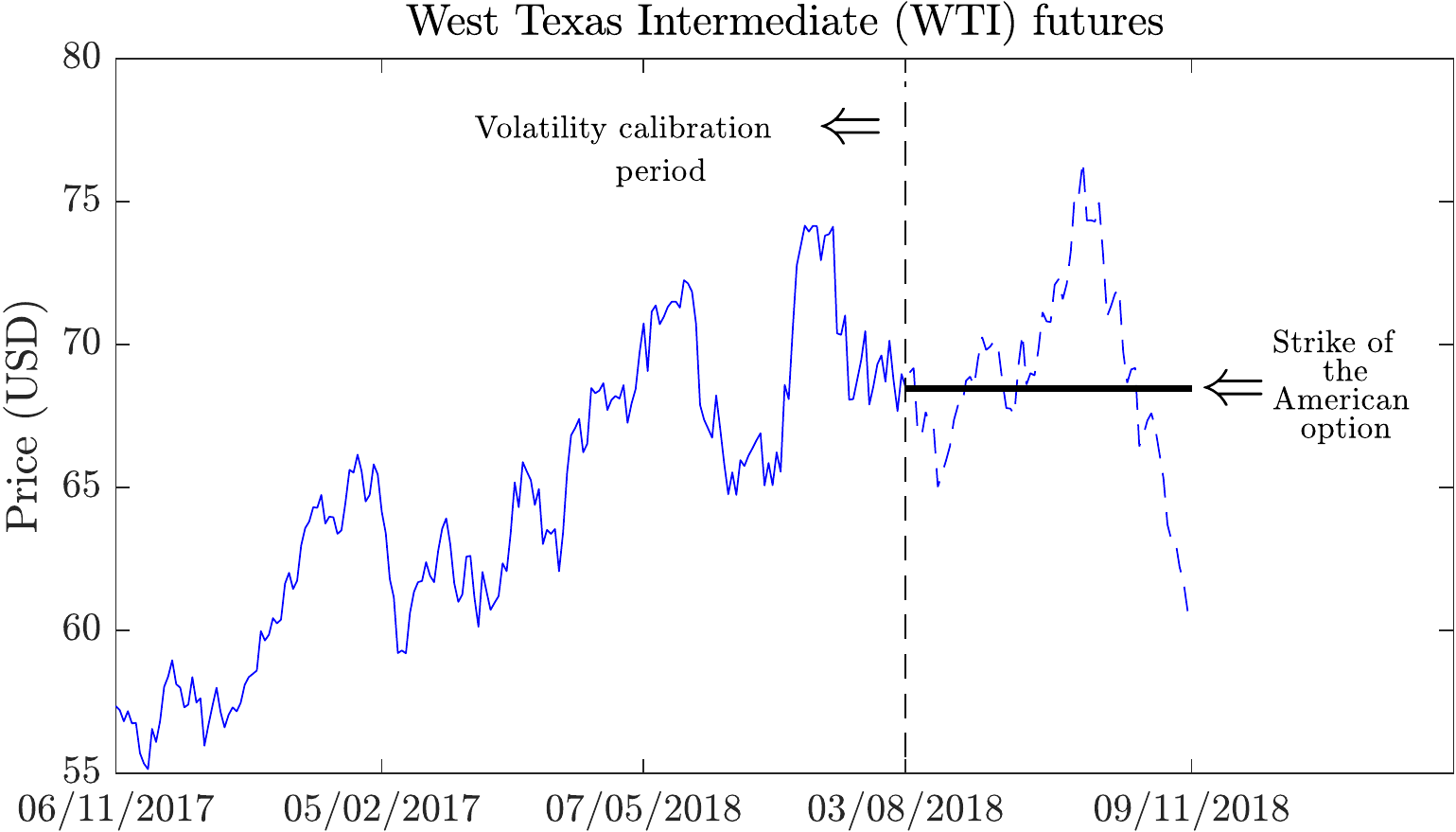}
			\caption{Daily closing prices of the West Texas Intermediate crude oil futures contracts. Data spans from 6.11.2017 to 9.11.2018. Vertical dashed line indicates the end of volatility calibration period. Option prices are calculated based on the volatility and the asset's price on 3.08.2018. }
			\label{fig:pi_kalibracjaWTI}
		\end{figure}

	We now focus on the potential application of $\pi$-options and compare the prices of American put and $\pi$-option on relative drawdown. We compare these particular instruments due to the fact that their values increase with the decrease of the underlying asset's price.
    As an exemplary environment for the options comparison we choose two time series containing
		daily closing prices of the Microsoft Corporation's stock (see Figure \ref{fig:pi_kalibracja}) as well as daily closing prices of the West Texas Intermediate (WTI) crude oil futures (see Figure \ref{fig:pi_kalibracjaWTI}). Both datasets are taken from \textit{www.finance.yahoo.com} and span approximately one year, from 6.11.2017 to 9.11.2018. We use the first $9$ months (from 6.11.2017 to 3.08.2018) to calibrate the historical volatility for both assets, which is one of the input parameters in our pricing algorithm.
		
	 Then, using the historical volatility, we compute prices of $\pi$ and American options (using assets' prices from 3.08.2018), both expiring $3$ months after the end of
		calibration period. Note that the parameters for the $\pi$-option on a relative drawdown are $a=-1$, $b=1$ and $K=1$.
Input parameters for calculation and estimated options prices for both assets are given in Table \ref{table:3} and Table \ref{table:4}.
		
Since the payoff of $\pi$-option on relative drawdown with $K=1$ is always less than $1$,
in order to compensate
		against the drop in underlying's price, we need a certain number of these contracts per each unit of stock in our portfolio.
		This number has to be equal to $M_0$. Note that in Table \ref{table:3} and Table \ref{table:4}, the real price of the single $\pi$-option on relative drawdown contract should be $0.0735$ for MSFT and $0.0949$ for WTI. However, in order to be able to compare the results to the American put values, we initially need to make the instruments pay the same amount in case of a price drop, therefore we multiply the price of single $\pi$-option on drawdown by $M_0$ ($110$ and $74$ for MSFT and WTI respectively). That is why in Table \ref{table:3} in Table \ref{table:4} the price of $\pi$-option equals $0.0735 \cdot 110 = 8.09$ for the stock and $0.094 \cdot 74 = 6.95$ for the oil futures contract.		

			\begin{table}[h!]
		\hspace*{3cm}
				\begin{tabular}{|c|c|c|c|}
					\hline
					\multicolumn{4}{|c|}{\bf{MSFT}} \\
					\hline
					\multicolumn{2}{|c|}{\bf{American put}}&\multicolumn{2}{|c|}{$\boldsymbol{\pi}$ \bf{on drawdown}}\\
					\hline
					Parameter & Value & Parameter & Value \\
					\hline
					$K$ & 108.13 & $M_0$ & 110.83\\
					\hline
					$S_0$ & \multicolumn{3}{|c|}{108.13} \\
					\hline
					$\sigma$ & \multicolumn{3}{|c|}{24.06 \%}\\
					\hline
					$r$ & \multicolumn{3}{|c|}{ 2.25\%} \\
					\hline
					$l$ & \multicolumn{3}{|c|}{100} \\
					\hline
					$T$ & \multicolumn{3}{|c|}{ 3} \\
					\hline
					\multicolumn{4}{|c|}{Option price}\\
					\hline
					\multicolumn{2}{|c|}{$\bf{\$5.47}$} & \multicolumn{2}{|c|}{$\bf{\$8.09}
$} \\
					\hline
				\end{tabular}
\vspace{0.5cm}	

				\caption{Input parameters for computation and estimated options' prices for the MSFT dataset.}
				 \label{table:3}
			\end{table}

			\begin{table}[h!]
		\hspace*{3cm}
				\begin{tabular}{|c|c|c|c|}
					\hline
					\multicolumn{4}{|c|}{\bf{WTI}} \\
					\hline
					\multicolumn{2}{|c|}{\bf{American put}} & \multicolumn{2}{|c|}{$\boldsymbol{\pi}$ \bf{on drawdown} }\\
					\hline
					Parameter & Value & Parameter & Value \\
					\hline
					$K$ & 68.49 & $M_0$ & 74.15\\
					\hline
					$S_0$ & \multicolumn{3}{|c|}{68.49} \\
					\hline
					$\sigma$ & \multicolumn{3}{|c|}{23.97 \%} \\
					\hline
					$r$ & \multicolumn{3}{|c|}{ 2.25\%} \\
					\hline
					$l$ & \multicolumn{3}{|c|}{100} \\
					\hline
					$T$ & \multicolumn{3}{|c|}{3} \\
					\hline
					\multicolumn{4}{|c|}{Option price}\\
					\hline
					\multicolumn{2}{|c|}{$\bf{\$3.07}$} & \multicolumn{2}{|c|}{$\bf{\$6.95}
$} \\
					\hline
				\end{tabular}
\vspace{0.5cm}	

				\caption{Input parameters for computation and estimated options' prices for the WTI dataset.}
				 \label{table:4}
			\end{table}
			
\begin{figure}[!h]
			\centering
			\includegraphics[width = .99\textwidth]{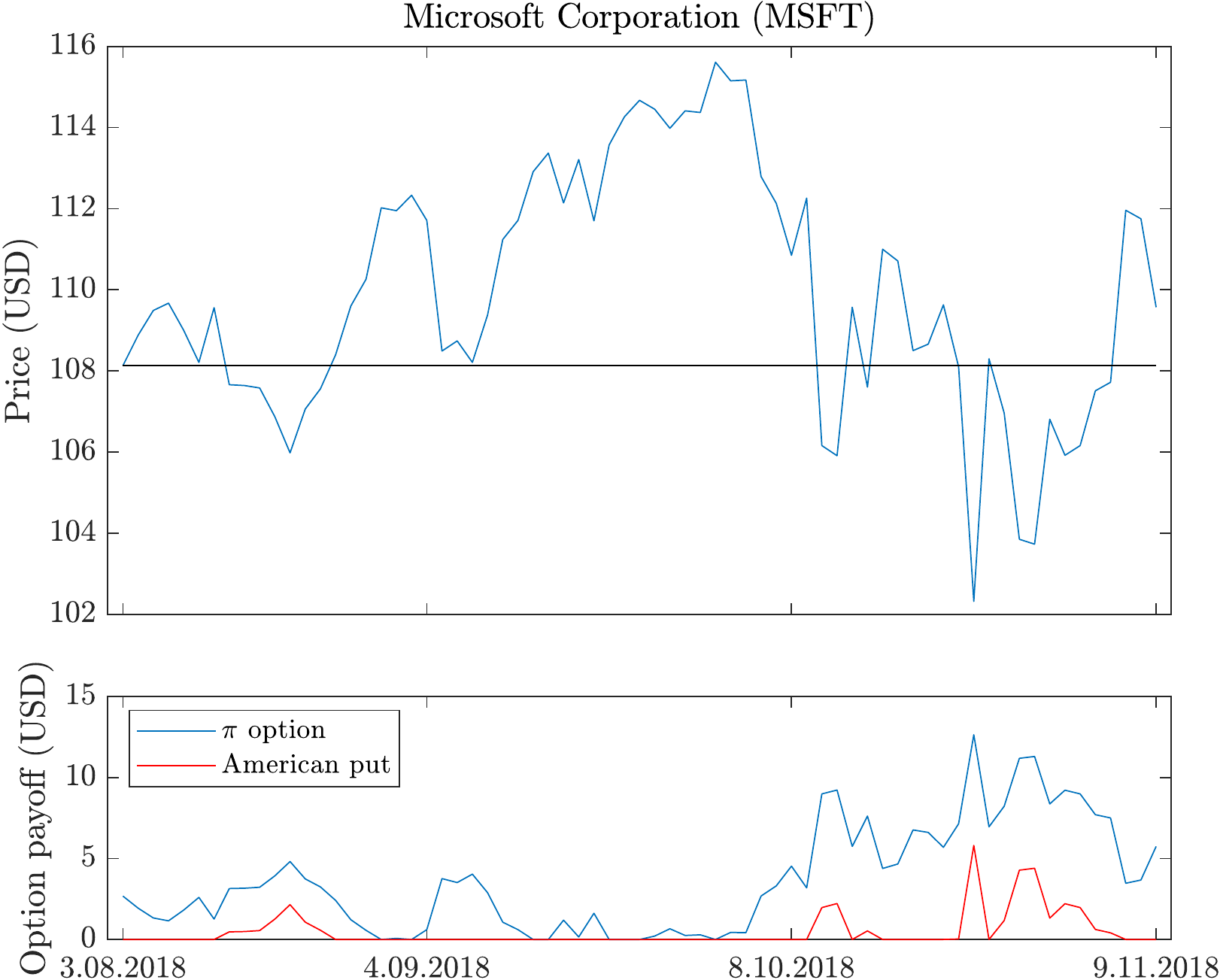}
			\caption{Microsoft Corporation stock closing prices (\textbf{top}) and the corresponding payoffs of $\pi$-option on relative drawdown and American put (\textbf{bottom}) with the parameters from Table \ref{table:3}.}
			\label{fig:pi_wyplaty}
		\end{figure}

\begin{figure}[!h]
			\centering
			\includegraphics[width = .99\textwidth]{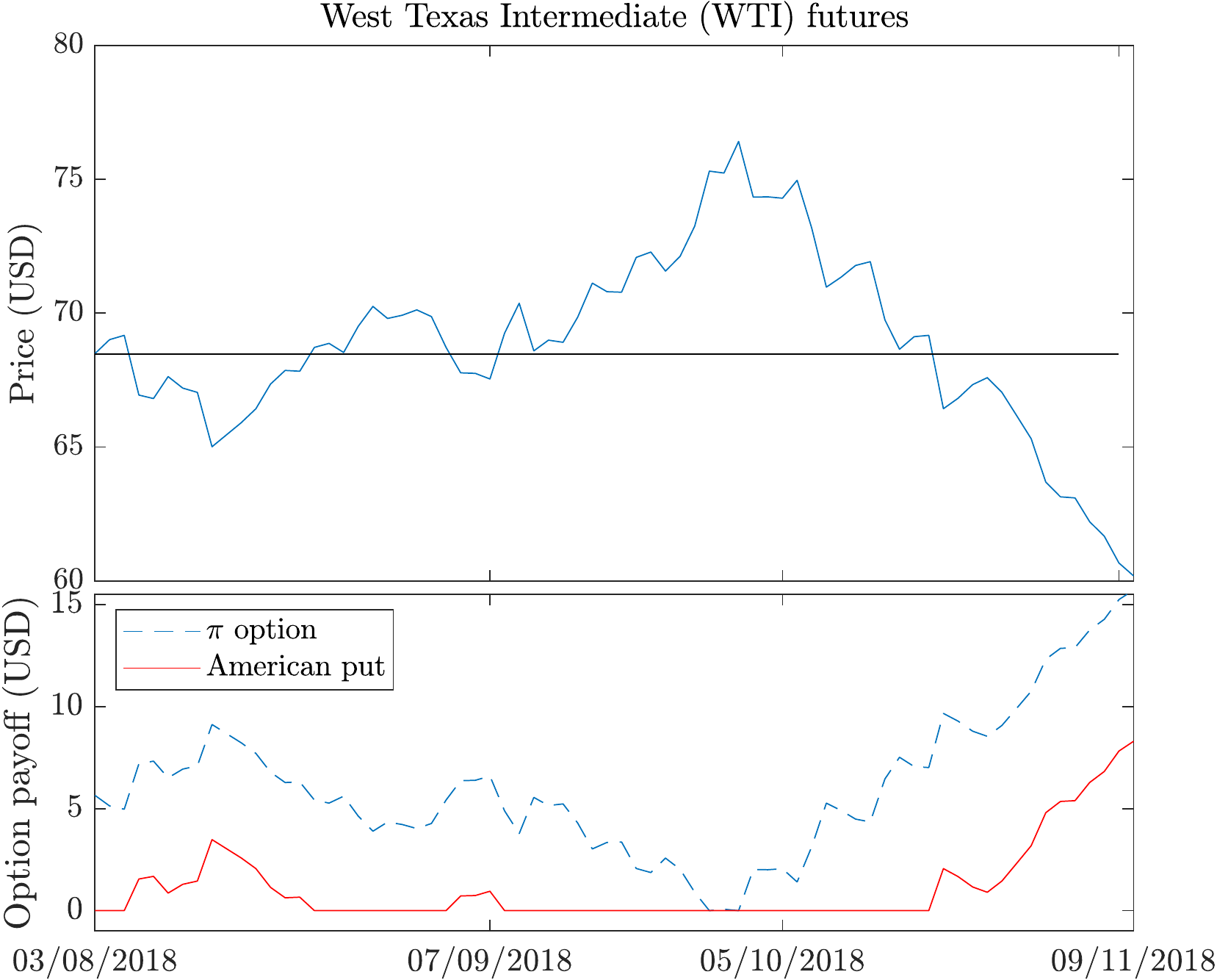}
			\caption{WTI crude oil futures contract closing prices (\textbf{top}) and the corresponding payoffs of $\pi$-option on relative drawdown and American put (\textbf{bottom}) with the parameters from Table \ref{table:4}.}
			\label{fig:pi_wyplatyWTI}
		\end{figure}

			
			It turns out that $\pi$-option is more expensive than vanilla put in case of both assets, which is not a surprise as it initially pays the amount equivalent to
		the present maximum drawdown.
	  However, since the difference in price between these instruments is rather significant, a question emerges whether there
		exists a situation in which purchasing $\pi$-option on relative drawdown is more profitable than buying a simple vanilla put.		
		In order to answer this question, let us focus on the dashed part of the Microsoft Corporation and WTI futures data from the beginning of this section.
		In Figures
		\ref{fig:pi_wyplaty} and \ref{fig:pi_wyplatyWTI} we show the amount each instrument would pay (on each day) throughout the whole $3$-month period until options' maturity.
			
		In order to display the difference more clearly, we construct two portfolios $V_{\text{American}}$ and $V_{\pi}$,
		both consisting of an underlying asset (a single Microsoft Corporation stock or a barrel of the WTI crude oil) and an option (American put and $\pi$-option on relative drawdown, respectively).
		We observe them at the end of the volatility calibration period. Assets' prices and options prices
		are taken from Table \ref{table:3} and Table \ref{table:4}. In Table \ref{table:5} and Table \ref{table:6} we show the initial net values of both $V_{\text{American}}$ and $V_{\pi}$ portfolios.
				
		\begin{table}[h!]
		\centering
				\begin{tabular}{|c|c|c|}
					\hline
					Portfolio & $\mathbf{V_{\text{\textbf{American}}}}$ & $\mathbf{V_{\boldsymbol{\pi}}}$ \\
					\hline
					Initial asset value & 108.13 & 108.13 \\
					\hline
					Option premium & -\$5.47 & -\$8.09 \\
					\hline
					Option initial payoff & \$0 & \$2.68 \\
					\hline
\textbf{Portfolio's net value} & \bf{\$102.66} & \bf{\$102.72} \\
					\hline
				\end{tabular}
\vspace{0.5cm}					
				\caption{Portfolios and their initial net values for the MSFT dataset.}
				 \label{table:5}
			\end{table}

					\begin{table}[h!]
		\centering
				\begin{tabular}{|c|c|c|}
					\hline
					Portfolio & $\mathbf{V_{\text{\textbf{American}}}}$ & $\mathbf{V_{\boldsymbol{\pi}}}$ \\
					\hline
					Initial asset value & 68.49 & 68.49 \\
					\hline
					Option premium & -\$3.07  & -\$6.95 \\
					\hline
					Option initial payoff & \$0 & \$5.66 \\
					\hline
\textbf{Portfolio's net value} & \bf{\$65.42} & \bf{\$67.20} \\
					\hline
				\end{tabular}
\vspace{0.5cm}					
				\caption{Portfolios and their initial net values for the WTI dataset.}
				 \label{table:6}
			\end{table}

\begin{figure}[!h]
			\centering
			\includegraphics[width = .99\textwidth]{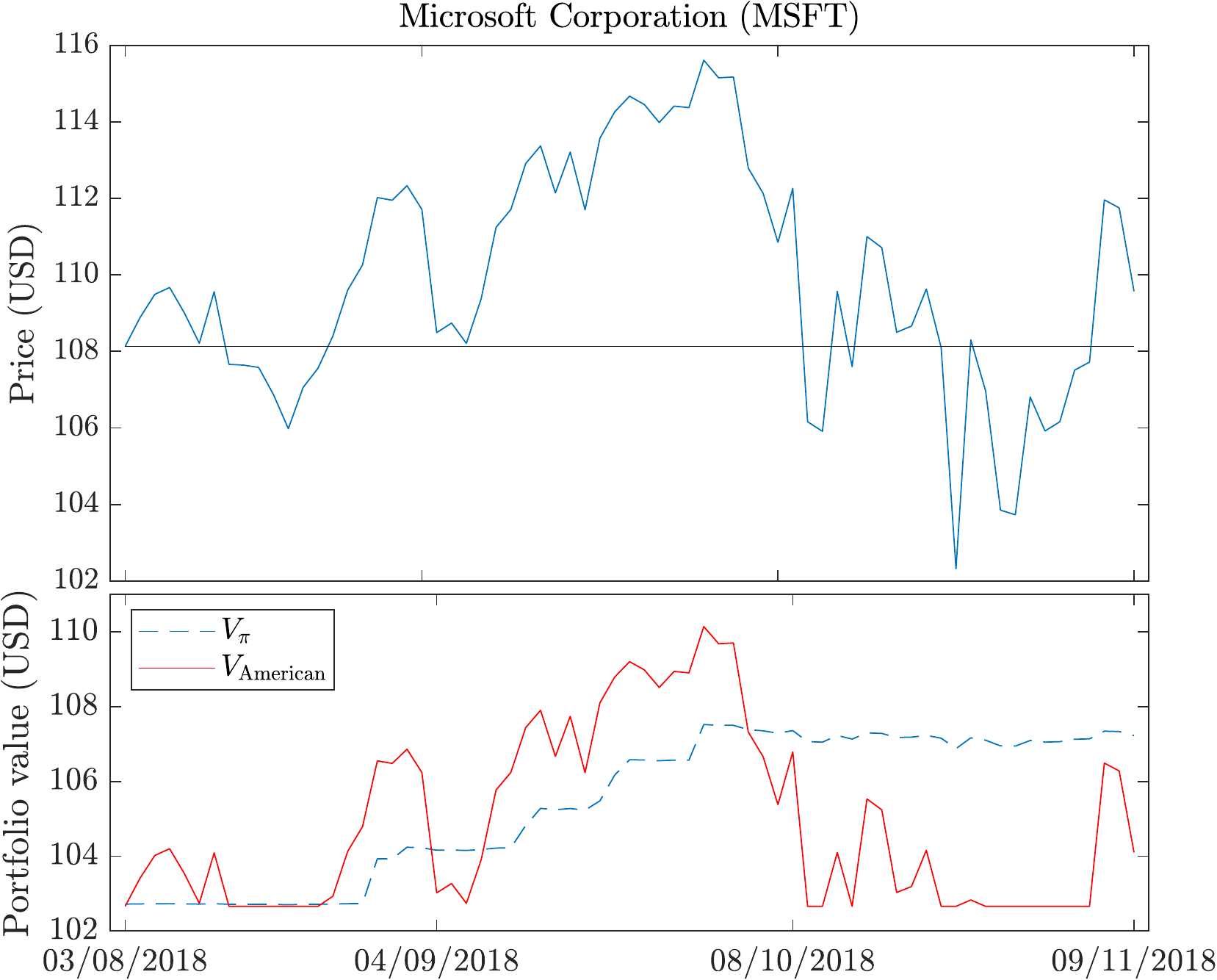}
			\caption{Microsoft Corporation stock closing prices (\textbf{top}) and payoffs of portfolios with parameters from Table \ref{table:3}
			(\textbf{bottom}).}
			\label{fig:pi_wyplaty_portfeli}
		\end{figure}

\begin{figure}[!h]
			\centering
			\includegraphics[width = .99\textwidth]{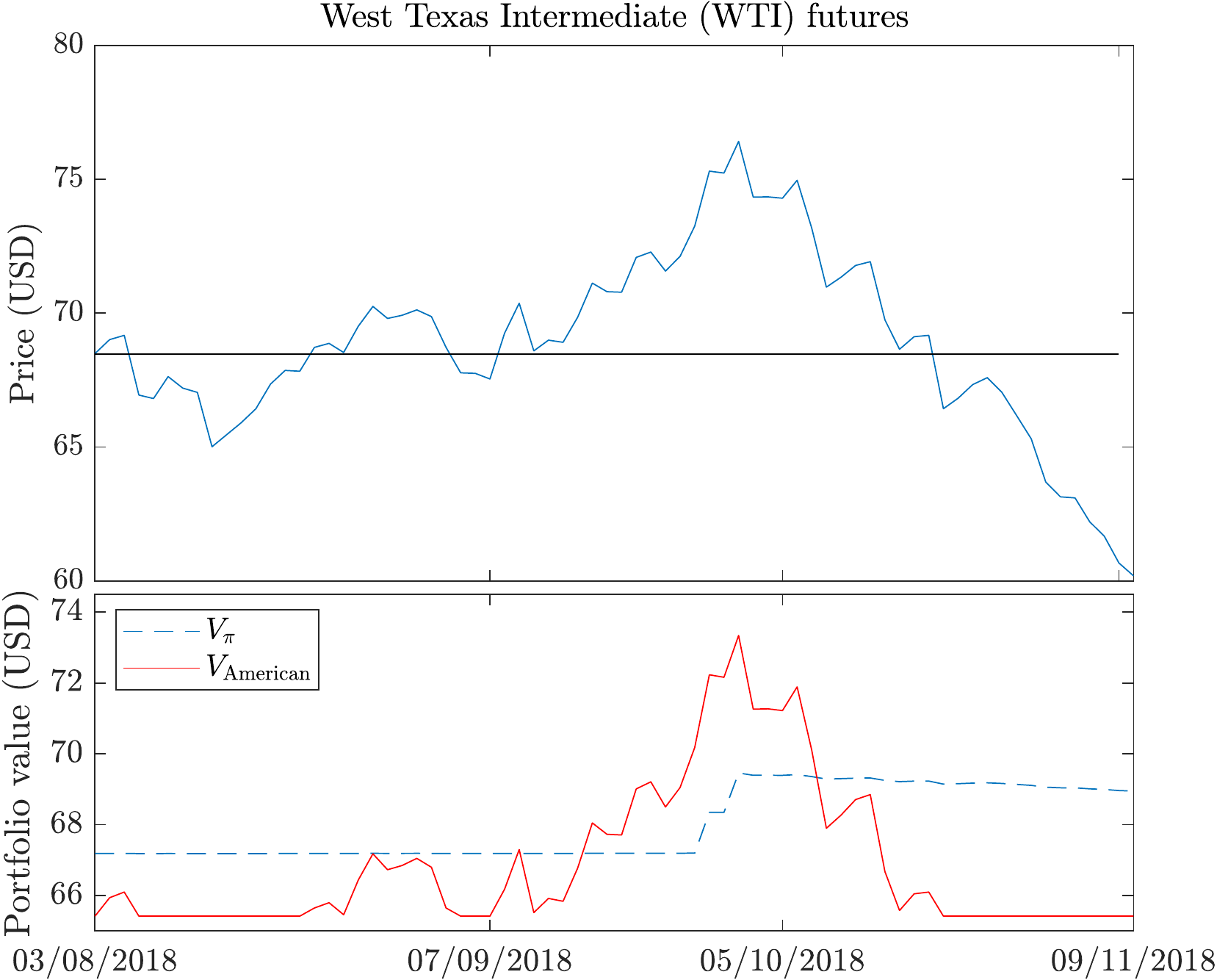}
			\caption{WTI crude oil futures contract closing prices (\textbf{top}) and payoffs of portfolios with parameters from Table \ref{table:4}
			(\textbf{bottom}).}
			\label{fig:pi_wyplaty_portfeliWTI}
		\end{figure}

Then we analyze the behaviour of the constructed portfolios, by calculating the net value of each portfolio for each day until options' maturity; see Figure
		\ref{fig:pi_wyplaty_portfeli} and Figure \ref{fig:pi_wyplaty_portfeliWTI}.
		
			Based on Figures \ref{fig:pi_wyplaty_portfeli} and \ref{fig:pi_wyplaty_portfeliWTI} we can observe that the maximum value of portfolio $V_{\text{American}}$ is greater than the one for $V_{\pi}$. Thus when focusing
			purely at the possible maximum profit over some period of time, then the portfolio containing American option performs better.
                   However, we can notice that $V_{\text{American}}$'s value over time is much more volatile compared
			to $V_{\pi}$ and it directly follows the behavior of underlying asset (it increases when asset's price rises and decreases in
			the opposite situation).
			The value $V_{\pi}$ of $\pi$-option portfolio is most of the time non-decreasing. Moreover, $V_{\pi}$ increases its value every time the asset's price
			reaches a new maximum and essentially does not decrease in case of any price drop. In other words, combining the underlying asset and $\pi$-option
			on drawdown allow us to lock in our profit whenever the price reaches its new maximum. \\
			This brings us to the conclusion that the purpose of using $\pi$-option on relative drawdown and an American put is completely
			different. Vanilla American option protects us from asset price drops and ensures us that the current worth of our portfolio will not be
			less than its initial value. Unfortunately, in this case our portfolio's value is more volatile and reflects the volatility of the underlying asset.
			This may result in bigger gains when comparing to the use of $\pi$-option on relative drawdown if the price of the underlying rises
			significantly and stays on that level until option's maturity. However, in case of a drop in asset price after the upswing,
			we do not benefit from the fact that the new maximum has been reached and thus the value of our portfolio decreases together with
			the price of the underlying asset.
			When looking at the value of $V_{\pi}$ over time one can notice that combining stock or a commodity and $\pi$-option on relative drawdown protects us
			against price drops as well but the volatility of our portfolio is reduced significantly. Additionally,
the contract allows us to benefit from the underlying's price upswings and locks in the profit every time
			new maximum is reached.

We have analyzed two datasets, MSFT and WTI, and the above analysis shows that the behaviour of a portfolio based on $\pi$-option is similar for various
choices of underlying assets.

		\section{Conclusions}\label{sec:con}

			In this paper we focus on the numerical pricing of the new derivative instrument - a $\pi$-option.
                  We adapted the Monte Carlo algorithm proposed by \cite{broadie_glasserman}
			to price this new option. 
			We focused on a specific parametrization of this option which we call the $\pi$-option on drawdown. We observed that this specific financial instrument
			is related to so-called relative maximum drawdown.
			We obtained prices of the $\pi$-option on relative drawdown for the Microsoft Corporation stock with
			different parameters in order to examine the influence of those parameters on option's premium.
			Our next step involved the analysis of two portfolios: first one based on a $\pi$-option on relative drawdown and the second one
			based on an American put. We used the Microsoft Corporation data as well as the West Texas Intermediate crude oil futures dataset.
			It turned out that the portfolios behave in a completely different manner.
			The value of the portfolio containing the American put was highly correlated with the underlying's price movements and thus
			had an unpredictable and volatile behavior.
			On the other hand, combining $\pi$-option on relative drawdown with the underlying asset not only
			ensures that the worth of the portfolio will not drop below the initial level, but it also allows us to take advantage of price upswings and to reduce the portfolio's volatility at the same time.
                  Similar analysis could be carried out for a geometric L\'evy process of asset price. One can also consider the regime-switching market.







\begin{thebibliography}{999}
	
\bibitem[Barraquand and Martineau (1995)]{1}
Barraquand, J. and Martineau, D. Numerical valuation of high dimensional multivariate American
securities. {\em J. Finan. Quant. Anal.} {\bf 1995}, {\em 30}, 383--405.

\bibitem[Brennan and Schwartz (1978)]{Schw1}
Brennan, M. J. and Schwartz E. S. Finite Difference Method and Jump Processes Arising in the Pricing
of Contingent Claims. {\em J. Finan. Quant. Anal.} {\bf 1978}, {\em 13}, 461--474.

\bibitem[Boyle (1977)]{2}
Boyle, P. Options: a Monte Carlo approach. {\em Journal of Financial Economics} {\bf 1977} {\em 4(3)}, 323--338.

\bibitem[Boyle et al. (1997)]{3}
Boyle, P., Broadie, M. and Glasserman, P. Monte Carlo Methods for Security Pricing. {\em Journal of Economic Dynamics and Control}
{\bf  1997}, {\em 21(8)}, 1267--1321.

\bibitem[Broadie and Detemple (1996)]{Broadie}
Broadie, M. and Detemple J. American Option Valuation, New Bounds, Approximations, and a Comparison of Existing Methods.
{\em  Review of Financial Studies} {\bf 1996}, {\em 9(4)}, 1211--1250.

\bibitem[Broadie et al. (1997)]{4}
Broadie, M., Glasserman, P. and Jain, G. Enhanced Monte Carlo estimates for American option prices.
{\em Journal of Derivatives} {\bf 1997}, {\em 5(1)}, 25--44.

\bibitem[Broadie and Glasserman (1997)]{broadie_glasserman}
Broadie, M. and Glasserman, P.  Pricing American-style securities using simulation.
{\em Journal of Economic Dynamics and Control} {\bf 1997}, {\em 21(8-9)}, 1323--1352.

\bibitem[Cl\'ement et al. (2002)]{5}
Cl\'ement, E., Lamberton, D. and Protter, P. An Analysis of a Least Squares Regression Method for American Option Pricing.
{\em Finance and Stochastics} {\bf 2002}, {\em 6(4)}, 449--471.

\bibitem[Caflisch (1998)]{6}
Caflisch, R. Monte Carlo and Quasi-Monte Carlo Methods. {\em Acta Numerica}, {\bf 1998}, {\em 7}, 1--49.

\bibitem[Carr and Faguet (1994)]{Carr}
Carr, P. and Faguet, D. Fast accurate valuation of American options.
{\em Working Paper, Cornell University} {\bf 1994}.

\bibitem[Carriere (1996)]{Carriere}
Carriere, J. F. Valuation of the early-exercise price for options
using simulations and nonparametric regression. {\em Insurance: Math. Economics} {\bf 1996}, {\em 19}, 19--30.

\bibitem[Christensen (2013)]{Chris}
Christensen, S. Contributions to the theory of optimal stopping. {\em Phd theses},
{\bf 2013}, see \url{https://d-nb.info/1020001488/34}.

\bibitem[Dyer and Jacob (1991)]{7}
Dyer, L. and Jacob, D. An Overview of
Fixed Income Option Models?
In {\em The Handbook of Fixed Income Securities}, edited by E J. Fabozzi,
742--73. Homewood, IU.: Business One Irwin,
{\bf 1991}.

\bibitem[Egami and Oryu (2017)]{Egami}
Egamim, M. and Oryu, T.
A direct solution method for pricing options involving the maximum process.
{\em Finance and Stochastics}, {\bf 2017}, {\em 21}, 967--993.

\bibitem[Geske and Shastri (1985)]{8}
Geske, R. and Shastri, K. Valuation by
Approximation: A Comparison of Alternative
Option Valuation Techniques. {\em Journal of Financial and Quantitative Analysis} {\bf 1985}, {\em 20(1)}, 45--71.

\bibitem[Glasserman (2004)]{9}
Glasserman, P. {\em Monte Carlo Methods in Financial Engineering.} Springer-Verlag, New York, {\bf 2004}.

\bibitem[Guo and Zervos (2010)]{zervos}
Guo, X. and Zervos, M. $\pi$ options. {\em Stochastic Processes and their Applications} {\bf 2010}, {\em 120(7)}, 1033--1059.

\bibitem[J\"{a}ckel (2002)]{10}
J\"{a}ckel, P.: {\em Monte Carlo Methods in Finance.} John Wiley, Chichester, U.K. {\bf 2002}.

\bibitem[Resenburg and Torrie (1993)]{11}
van Resenburg, E.J. and Torrie, G.M. Estimation of multidimensional integrals: is Monte Carlo the best method?
{\em Journal of Physics A: Mathematics and General} {\bf 1993}, {\em 26(4)}.

\bibitem[Joy et al. (1996)]{12}
Joy, C., Boyle, P.P. and Tan, K.S. Quasi-Monte Carlo methods in numerical finance. {\em Management Science} {\bf 1996},
{\em 42(6)}.

\bibitem[Longstaff and Schwartz (2001)]{11b}
Longstaff, F.A. and Schwartz, E.S. Valuing American options by simulation:
A simple least-squares approach. {\em Review of Financial Studies} {\bf 2001}, 113--147.

\bibitem[Niederreiter (1992)]{13}
Niederreiter, H. {\em Random Number Generation and Quasi-Monte Carlo Methods.} Society
for Industrial and Applied Mathematics, Austria, {\bf 1992}.

\bibitem[Raymar and Zwecher (1997)]{13b}
Raymar, S. and Zwecher, M. A Monte Carlo valuation of American call options on the maximum
of several stocks. {\em J. Derivatives} {\bf 1997}, {\em 5}, 7--23.

\bibitem[Rogers (2002)]{14}
Rogers, L.C.G. Monte Carlo Valuation of American Options. {\em Mathematical Finance} {\bf 2002}, {\em 12(3)}, 271--286.

\bibitem[Schwartz (1977)]{Schw2}
Schwartz, E. The Valuation of Warrants: Implementing a New Approach.
{\em Journal of Financial Economics} {\bf 1977}, {\em 4}, 79--93.

\bibitem[Stentoft (2004a)]{Sten1}
Stentoft, L. Convergence of the Least Squares Monte Carlo Approach to American Option Valuation.
{\em Management Science} {\bf 2004}, {\em 50(9)}, 1193--1203.

\bibitem[Stentoft (2004b)]{Sten2}
Stentoft, L. Assessing the Least Squares Monte Carlo Approach to American Option Valuation.
{\em Review of Derivatives Research} {\bf 2004}, {\em 7(2)}, 129--168.

\bibitem[Tilley (1993)]{14b}
Tilley, J. A. Valuing American Options in a Path simulation model. {\em Transactions of the Society of Actuaries} {\bf 1993},
{\em 15}, 499--550.

\bibitem[Tsitsiklis and van Roy (1999)]{15}
Tsitsiklis, J.N.  and van Roy, B. Optimal stopping of Markov processes: Hilbert space
theory, approximation algorithms, and an application to pricing high-dimensional financial derivatives.
{\em IEEE Transactions on Automatic Control} {\bf 1999}, {\em 44(10)}, 1840--1851.

\bibitem[Tsitsiklis and van Roy (2001)]{16}
Tsitsiklis, J.N.  and van Roy, B. Regression methods for pricing complex American-style
options. {\em IEEE Transactions on Neural Networks} {\bf 2001}, {\em 12(4)}, 694--703.

\bibitem[Villani (2010)]{17}
Villani, G. A Monte Carlo approach to value exchange options
using a single stochastic factor.
In
{\em Mathematical and Statistical Methods for Actuarial Sciences and Finance}, edited by
Ballester, Ferrer, Corazza, Pizzi, Springer-Verlag {\bf 2010}, 305--331.

\bibitem[Zhao (2018)]{Zhao}
Zhao, J. American Option Valuation Methods.
{\em International Journal of Economics and Finance} {\bf 2018}, {\em 10(5)}.

\end{thebibliography}
\end{document}